\documentclass{article}
\usepackage[hidelinks]{hyperref}
\usepackage{amsmath}
\usepackage{parskip} 
\usepackage{natbib}
\usepackage{mathrsfs}
\usepackage{pdfpages}
\usepackage{bm}
\usepackage{booktabs}
\usepackage{graphics}
\usepackage{multicol}
\usepackage[bf]{caption}
\usepackage{color}
\usepackage{amssymb}
\usepackage{multirow}
\usepackage{booktabs}
\usepackage{subcaption}

\usepackage{comment}
\usepackage{amsthm}

\usepackage[margin=1in]{geometry}
\usepackage{floatrow}
\newfloatcommand{capbtabbox}{table}[][\FBwidth]
\usepackage{rotating}
\usepackage[titletoc]{appendix}
\usepackage[graphicx]{realboxes}
\usepackage{blindtext,titlefoot}
\usepackage{setspace}

\title{\textbf{On constraints imposed on a transversely isotropic elasticity tensor }}

\author{Filip P. Adamus\footnote{%
Department of Earth Sciences, Memorial University of Newfoundland,
{\tt adamusfp@gmail.com}}\, and\,
Izabela Kudela\footnote{%
Department of Earth Sciences, Memorial University of Newfoundland,
{\tt izabelakudela@gmail.com}}\, }
\date{}

\begin{document}
\maketitle
\begin{abstract}
We discuss several physical constraints imposed on elasticity parameters of a transversely isotropic (TI) tensor.
There are three types of restrictions we investigate; a fundamental one of stability conditions, and two additional ones, commonly considered in seismology. 
The first commonly considered restriction comes from an assumption of a wave with a greater speed in the horizontal than vertical direction.
The second constitute the assumption that quasi-P wave is faster than quasi-S waves.
We show several numerical examples to examine how these restrictions affect a TI tensor with known values of certain elasticity constants that could be acquired from the vertical or horizontal measurements. 
\end{abstract}
\section{Introduction}
In this paper, we consider a transversely isotropic (TI) elasticity tensor with a symmetry axis that coincides with a vertical $x_3$-axis of a coordinate system that reffers to depth.
Such an elasticity tensor might be a good analogy to the mechanical properties of laminated rocks, for instance, shales or schists.
A TI tensor has five independent elasticity parameters; estimating their values via seismic measurements is not an easy task.
For instance, one may perform vertical seismic profiling at offsets close to zero, to obtain vertical speeds of quasi-P and quasi-S waves.
This way, five unknown parameters may be reduced to only three.
Another approach to reduce the number of unknown constants is to perform horizontal profiling in a region of a desired medium. 
By doing so, we get horizontal speeds of quasi-P and quasi-S waves, and we may obtain values of additional two elasticity parameters.
Most often we do not have an opportunity of knowing the values of all five parameters, but we still might want to use a TI tensor to better describe the properties of a medium.
Therefore, it is useful to restrict the range of the possible values of the remaining elasticity parameters that are unknown.
In other words, putting several constraints on these parameters allows us to better predict or estimate their values.

In Section~\ref{sec:two}, we describe a TI tensor and we discuss several constraints imposed on it.
In Section~\ref{sec:three}, we show numerical examples based on four distinct TI tensors.
In each case, the values of a different set of the elasticity parameters are known. 
Their known values might correspond to the real cases of acquiring the data from, respectively: vertical and horizontal seismic measurements of quasi-P wave, vertical measurements of quasi-P and quasi-S waves, horizontal measurements of quasi-P and quasi-S waves, and vertical and horizontal measurements of quasi-P and quasi-S waves.
In the last section, we discuss the obtained results. 
\section{Restrictions on elasticity parameters} \label{sec:two}
\subsection{Fundamental constraints}\label{fund}
Fundamental physical restrictions imposed on a elasticity tensor are the stability conditions.
They express the fact that it is necessary to expend energy to deform a material~\citep[e.g.][Section 4.3]{Red}. 
These conditions mean that every elasticity tensor must be positive-definite, wherein a tensor is positive-definite if and only if all eigenvalues of its symmetric-matrix representation are positive.

A TI tensor, whose rotation symmetry axis coincides with $x_3$-axis, may be expressed in a matrix form using Kelvin notation as,
\begin{equation}
C^{\rm TI}=
\left[
\begin{array}{cccccc}
c_{1111} & c_{1122} & c_{1133} & 0 & 0 & 0\\
c_{1122} & c_{1111} & c_{1133} & 0 & 0 & 0\\
c_{1133} & c_{1133} & c_{3333} & 0 & 0 & 0\\
0 & 0 & 0 & 2c_{2323} & 0 & 0\\
0 & 0 & 0 & 0 & 2c_{2323} & 0\\
0 & 0 & 0 & 0 & 0 & 2c_{1212}
\label{ti}
\end{array}
\right]
\,,
\end{equation}
where $c_{1122}=c_{1111}-2c_{1212}$. 
Hence, the elasticity tensor from expression~(\ref{ti}) has five independent parameters.
Its eigenvalues are
\begin{equation*}\label{eig1}
\lambda_1= \lambda_2=2c_{1212} \,,\qquad \lambda_3=\lambda_4=2c_{2323} \,,
\end{equation*}
\begin{equation*}
\lambda_5=\frac{1}{2}\left(2c_{1111}-2c_{1212}+c_{3333}+\sqrt{(2c_{1111}-2c_{1212}-c_{3333})^2+8(c_{1133})^2}\,\right) \,,
\end{equation*}
\begin{equation*}\label{eig5}
\lambda_6=\frac{1}{2}\left(2c_{1111}-2c_{1212}+c_{3333}-\sqrt{(2c_{1111}-2c_{1212}-c_{3333})^2+8(c_{1133})^2}\,\right) \,.
\end{equation*}
To satisfy the stability conditions the eigenvalues must be positive, after algebraic manipulation, we obtain 
\begin{align}\label{stab}
\begin{array}{c}
c_{1212}>0\,,\quad
c_{2323}>0\,,\quad
c_{3333}>0\,,\quad
c_{1111}-c_{1212}>0\,,\quad \\
\hphantom{x}\\
\left(c_{1111}-c_{1212}\right)c_{3333}>\left(c_{1133}\right)^2\,,
\end{array}
\end{align}
which are the fundamental constraints imposed on a TI tensor.  
\subsection{Common constraints}\label{common}
Let us consider the speeds of quasi-P (qP), quasi-transverse (SV), and transverse (SH) waves in a TI medium in both vertical (vrt) and horizontal (hor) directions of propagation.
\begin{equation*}
V_{qP}({\rm vrt})=\sqrt{\frac{c_{3333}}{\rho}}\,,\qquad
V_{SV}({\rm vrt})=\sqrt{\frac{c_{2323}}{\rho}}\,,\qquad
V_{SH}({\rm vrt})=\sqrt{\frac{c_{2323}}{\rho}}\,,
\end{equation*}
\begin{equation*}
V_{qP}({\rm hor})=\sqrt{\frac{c_{1111}}{\rho}}\,,\qquad
V_{SV}({\rm hor})=\sqrt{\frac{c_{2323}}{\rho}}\,,\qquad
V_{SH}({\rm hor})=\sqrt{\frac{c_{1212}}{\rho}}\,,
\end{equation*}
where $\rho$ denotes density.
Commonly, seismic waves propagate faster in the horizontal direction than vertical one.
Thus, we may introduce two constraints,
\begin{equation}\label{con1}
c_{1111}>c_{3333}\qquad {\mathrm {and}}\qquad c_{1212}>c_{2323}\,,
\end{equation}
which come from the speeds of qP and SH waves, respectively.
Another assumption we can make is that qP wave propagates faster than SV or SH wave, if they propagate in the horizontal or in the vertical direction.
Hence, we obtain 
\begin{equation}\label{con2}
c_{3333}>c_{2323}\,,\qquad c_{1111}>c_{2323}\,,\qquad {\mathrm {and}}\qquad c_{1111}>c_{1212}\,.
\end{equation}
We notice that the last constraint, coming from the assumption that $V_{qP}({\mathrm {hor}})>V_{SH}({\mathrm {hor}})$, is included in the stability conditions from expression~(\ref{stab}).

Some of the constraints from both expressions~(\ref{con1}) and~(\ref{con2}) are related to each other. 
To show it, let us use the relation $c_{1122}=c_{1111}-2c_{1212}$, and rewrite the last inequality of expression~(\ref{con2}), $c_{1111}>c_{1212}$, as,
\begin{equation}\label{ineq}
c_{1122}+c_{1212}>0\,.
\end{equation}
Then, we rewrite middle inequality of expression~(\ref{con2}), $c_{1111}>c_{2323}$, as,
\begin{equation*}
c_{1122}+c_{1212}+c_{1212}>c_{2323}\,,
\end{equation*}
and from inequality~(\ref{ineq}), we see that the sum of the first two terms on the left-hand side is a positive quantity and two other terms represent the last inequality of expression~(\ref{con1}).
In other words, a constraint $c_{1111}>c_{2323} \Leftrightarrow |a|+c_{1212}>c_{2323}$, is obvious from constraint $c_{1212}>c_{2323}$.

To summarize, apart from the fundamental constraints, we may add three independent and commonly considered assumptions, namely,
\begin{equation}\label{eq:common}
c_{1111}-c_{3333}>0\,,\qquad c_{1212}-c_{2323}>0\,,\qquad {\mathrm {and}}\qquad c_{3333}-c_{2323}>0\,.
\end{equation}
\section{Numerical examples}\label{sec:three}
In this section, we study the fundamental and common constraints---shown respectively in Sections~\ref{fund} and~\ref{common}---imposed on four examples of a TI medium.
In the first case, we examine a TI tensor with the known values of $c_{1111}$ and $c_{3333}$.
In the second case, the values of $c_{2323}$ and $c_{3333}$ are known, in the third, the values of $c_{1111}, c_{2323}$ and $c_{1212}$, and in the last one, the values of $c_{1111}, c_{3333}, c_{2323}$ and $c_{1212}$.
The values of the elasticity parameters are based on the Green-river shale, as shown by \citet{Thomsen} and examplified by~\citet[Exercise 9.3]{Red}.
\subsection{TI tensor based on qP-wave information} \label{sec:caseone}
\begin{figure}[H]
\centering
\begin{subfigure}{.45\textwidth}
  \centering
   \includegraphics[scale=0.38]{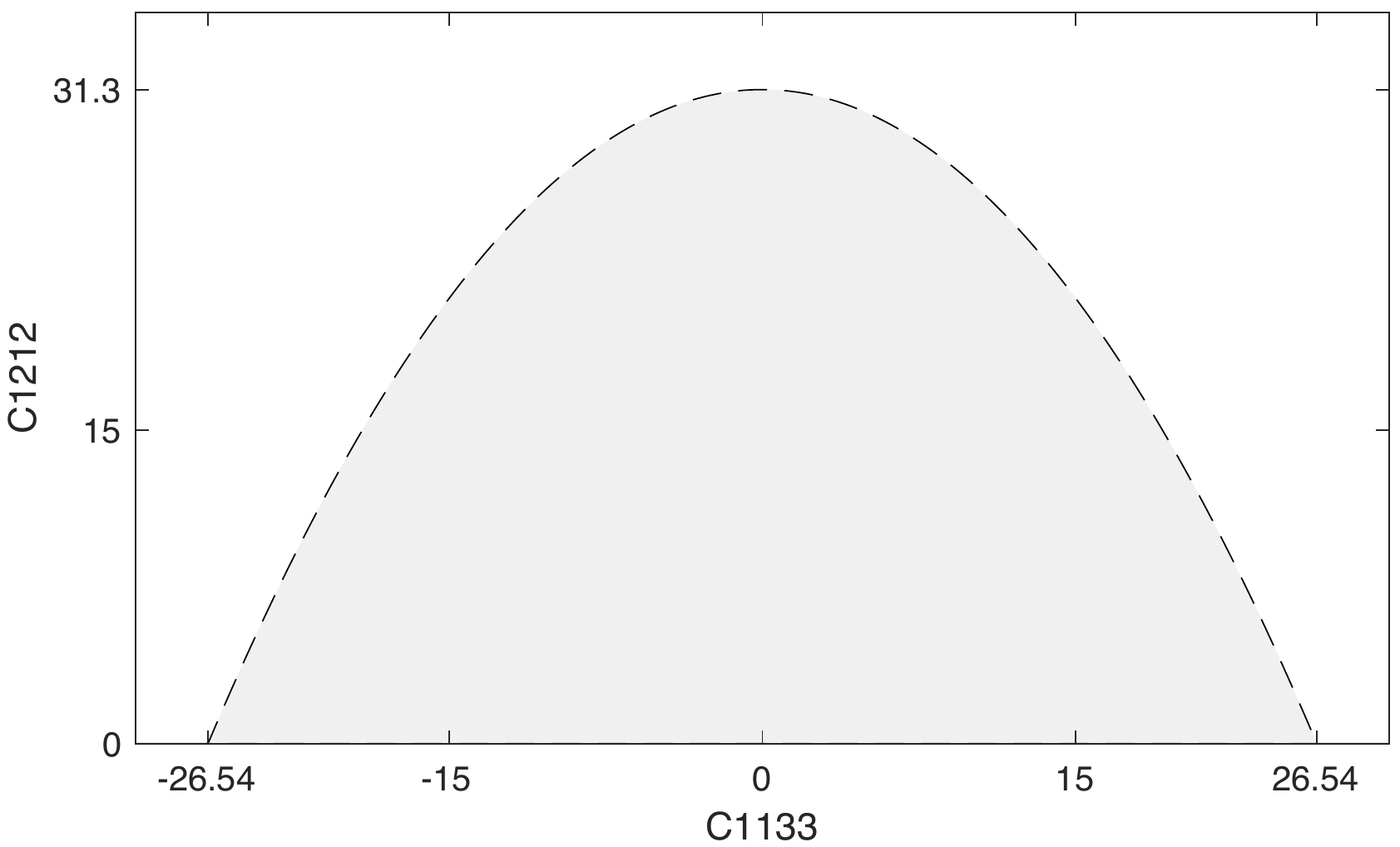}
  \caption{\footnotesize{no influence of common constraints}}
  \label{fig:1a}
\end{subfigure}%
\qquad
\begin{subfigure}{.45\textwidth}
  \centering
   \includegraphics[scale=0.38]{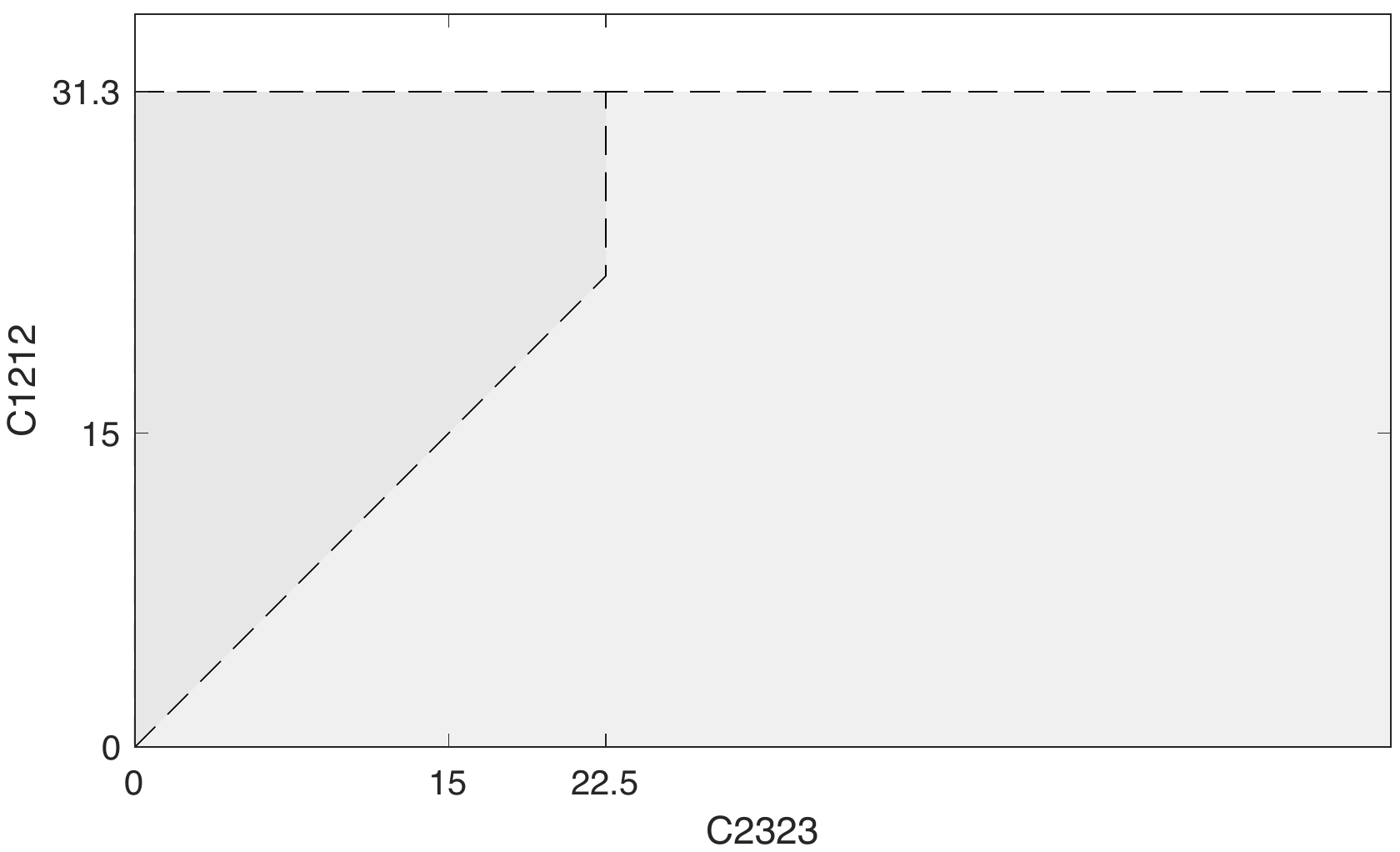}
  \caption{\footnotesize{common constraints restrict the area}}
  \label{fig:1b}
\end{subfigure}%
\caption{\small{The area of all possible values of unknown elasticity parameters restricted by the fundamental constraints is shown by the light grey colour. The dark grey area is the intersection of both restricted areas that come from the common and from the fundamental constraints. }}
\label{fig:one}
\end{figure}

Let us consider a TI tensor with given two elasticity parameters, $c_{1111}= 31.3\,[\rm{GPa}]$ and $c_{3333}=22.5\,[\rm{GPa}]$, where GPa are gigapascals.
This case may be relevant to the studies of a TI medium based on qP-wave information. 
Its eigenvalues are
$$\lambda_1=\lambda_2=2c_{1212}\,,\qquad \lambda_3=\lambda_4=2c_{2323}\,,$$
 \begin{equation*}
\lambda_5=42.55-c_{1212}+\frac{1}{2}\sqrt{(2c_{1212}-40.1)^2+8(c_{1133})^2}\,,
\end{equation*} 
  \begin{equation*}
\lambda_6=42.55-c_{1212}-\frac{1}{2}\sqrt{(2c_{1212}-40.1)^2+8(c_{1133})^2}\,,
\end{equation*} 
and its corresponding fundamental constraints from expression~(\ref{stab}) are
\begin{equation*}
0<c_{1212}<31.3\,,\qquad c_{2323}>0\,,\qquad 22.5>0\,,\qquad c_{1212}<31.3-\frac{(c_{1133})^2}{22.5}\,,
\end{equation*}
where $c_{1133}\in (-26.54\,,\,26.54)$.
Imposing the common constraints from expression~(\ref{eq:common}), we obtain
\begin{equation*}
8.8>0\,,\qquad c_{2323}<22.5\,,\qquad {\mathrm {and}}\qquad c_{1212}-c_{2323}>0\,.
\end{equation*}
The effect of imposing the common constraints, as an additional restriction to the fundamental one, is illustrated in Figures~\ref{fig:1a} and~\ref{fig:1b}.
\subsection{TI tensor based on information along symmetry axis}\label{sec:casetwo}
\begin{figure}[!htbp]
\centering
   \includegraphics[scale=0.38]{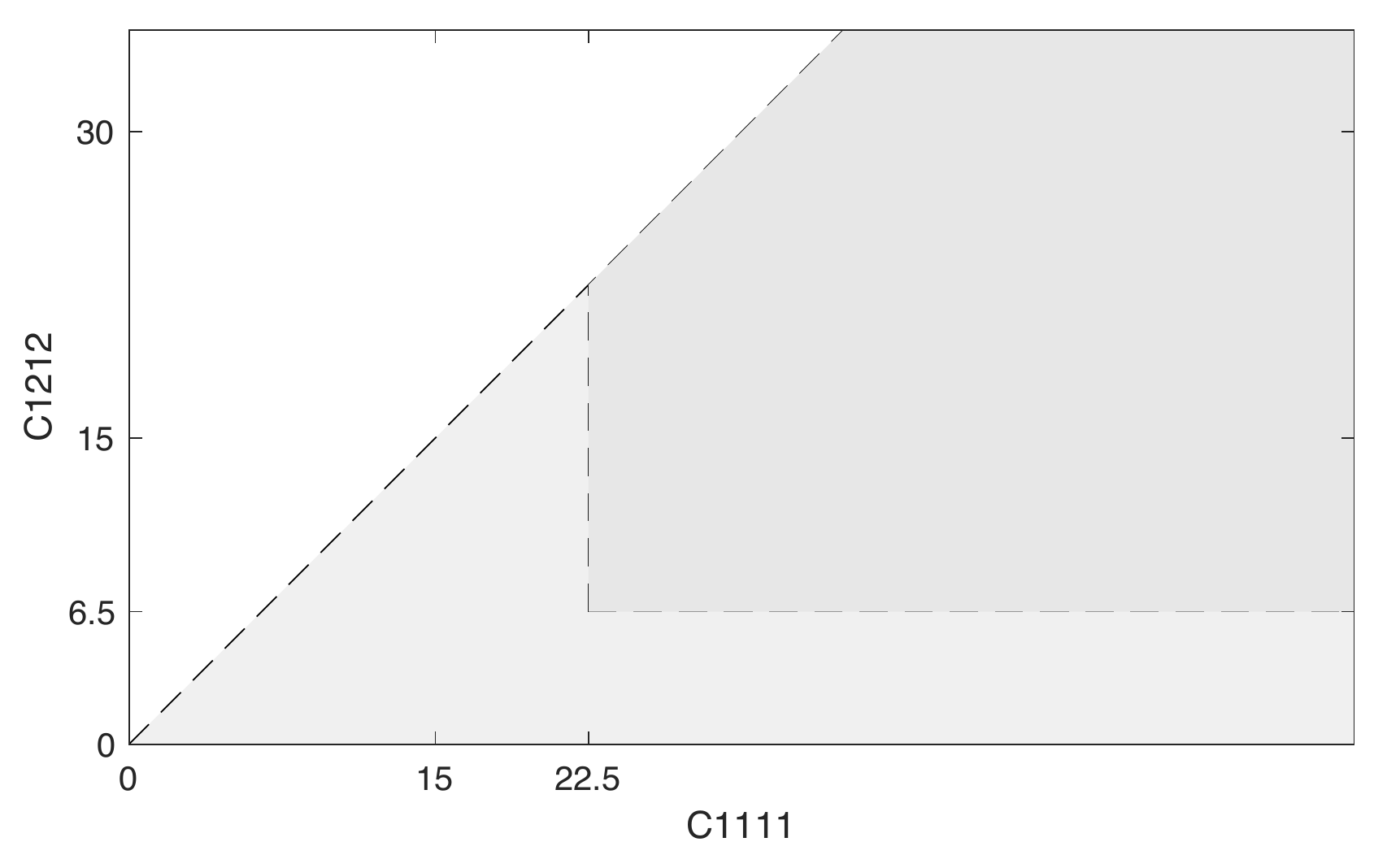}
  \caption{\small{The area of all possible values of unknown elasticity parameters restricted by the fundamental constraints is shown by the light grey colour. The dark grey area is the intersection of both restricted areas that come from the common and from the fundamental constraints. }}
\label{fig:two}
\end{figure}

Let us consider a TI tensor with given two elasticity parameters, $c_{2323}= 6.5\,[\rm{GPa}]$ and $c_{3333}=22.5\,[\rm{GPa}]$.
This case may be relevant to the studies of a TI medium based on measurements of qP and quasi-S waves along the symmetry axis. 
Its eigenvalues are
$$\lambda_1=\lambda_2=13\,,\qquad \lambda_3=\lambda_4=2c_{1212}\,,$$
 \begin{equation*}
\lambda_5=c_{1111}-c_{1212}+11.25+\frac{1}{2}\sqrt{(2c_{1212}-2c_{1111}+22.5)^2+8(c_{1133})^2}
\end{equation*} 
 \begin{equation*}
\lambda_6=c_{1111}-c_{1212}+11.25-\frac{1}{2}\sqrt{(2c_{1212}-2c_{1111}+22.5)^2+8(c_{1133})^2}
\end{equation*}
and its corresponding fundamental constraints are
\begin{equation*}
c_{1212}>0\,,\qquad 6.5>0\,,\qquad 22.5>0\,,\qquad c_{1111}-c_{1212}>0\,,\qquad (c_{1111}-c_{1212})22.5>(c_{1133})^2\,.
\end{equation*}
Imposing the common constraints from expression~(\ref{eq:common}), we obtain
\begin{equation*}
c_{1111}>22.5\,,\qquad 16>0\,,\qquad {\mathrm {and}}\qquad c_{1212}>6.5\,.
\end{equation*}
The effect of imposing the common constraints, as an additional restriction to the fundamental one, is illustrated in Figure~\ref{fig:two}.
\subsection{TI tensor based on information along horizontal axis}\label{sec:casethree}
\begin{figure}[!htbp]
\centering
   \includegraphics[scale=0.38]{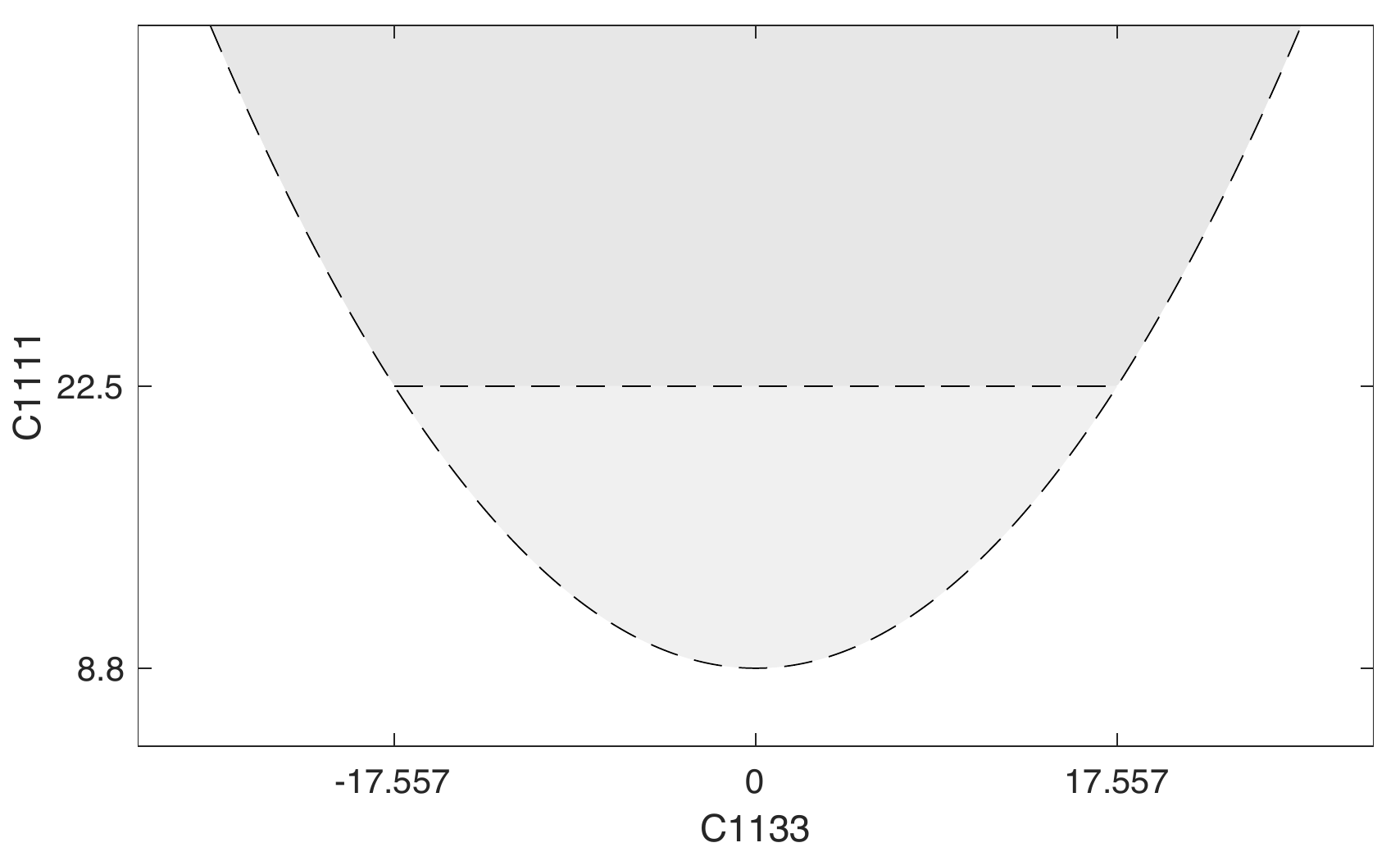}
  \caption{\small{The area of all possible values of unknown elasticity parameters restricted by the fundamental constraints is shown by the light grey colour. The dark grey area is the intersection of both restricted areas that come from the common and from the fundamental constraints. }}
\label{fig:three}
\end{figure}

Let us consider a TI tensor with given three elasticity parameters, $c_{1212}=8.8\,[\rm{GPa}],\,c_{2323}= 6.5\,[\rm{GPa}]$ and $c_{3333}=22.5\,[\rm{GPa}]$.
This case may be relevant to the studies of a TI medium based on measurements of qP, SV, and SH waves along the horizontal axis. 
Its eigenvalues are
$$\lambda_1=\lambda_2=13\,,\qquad \lambda_3=\lambda_4=17.6\,,$$
 \begin{equation*}
\lambda_5=c_{1111}+2.45+\frac{1}{2}\sqrt{(40.1-2c_{1111})^2+8(c_{1133})^2}\,,
\end{equation*} 
 \begin{equation*}
\lambda_6=c_{1111}+2.45-\frac{1}{2}\sqrt{(40.1-2c_{1111})^2+8(c_{1133})^2}\,,
\end{equation*}
and its corresponding fundamental constraints are
\begin{equation*}
8.8>0\,,\qquad 6.5>0\,,\qquad 22.5>0\,,\qquad c_{1111}>8.8\,,\qquad (c_{1111}-8.8)22.5>(c_{1133})^2\,.
\end{equation*}
Imposing the common constraints from expression~(\ref{eq:common}), we obtain
\begin{equation*}
c_{1111}>22.5\,,\qquad 16>0\,,\qquad {\mathrm {and}}\qquad 2.3>0\,.
\end{equation*}
The effect of imposing the common constraints, as an additional restriction to the fundamental one, is illustrated in Figure~\ref{fig:three}.

\subsection{TI tensor based on information along symmetry and horizontal axes}\label{sec:casefour}
Let us consider a TI tensor with given four elasticity parameters, $c_{1212}=8.8\,[\rm{GPa}],\,c_{2323}= 6.5\,[\rm{GPa}],\,c_{1111}=31.3\,[\rm{GPa}]$ and $c_{3333}=22.5\,[\rm{GPa}]$.
This case may be relevant to the studies of a TI medium based on measurements of qP, SV, and SH waves along both the symmetry and horizontal axes. 
Its eigenvalues are
$$\lambda_1=\lambda_2=13\,,\qquad \lambda_3=\lambda_4=17.6\,,$$
 \begin{equation*}
\lambda_5=33.75+\frac{1}{2}\sqrt{8(c_{1133})^2+506.25}\,,
\end{equation*} 
 \begin{equation*}
\lambda_6=33.75-\frac{1}{2}\sqrt{8(c_{1133})^2+506.25}\,,
\end{equation*}
and its corresponding fundamental constraints are
\begin{equation*}
8.8>0\,,\qquad 6.5>0\,,\qquad 22.5>0\,,\qquad 22.5>0\,,\qquad -22.5>c_{1133}>22.5\,.
\end{equation*}
Imposing the common constraints from expression~(\ref{eq:common}), we obtain
\begin{equation*}
8.8>0\,,\qquad 16>0\,,\qquad {\mathrm {and}}\qquad 2.3>0\,.
\end{equation*}
The common constraints do not additionally limit the possible values of $c_{1133}$.
In this case, only the stability conditions are valuable.

\section{Discussion}\label{sec:four}
In three out of four examples of a TI tensor shown in the previous section, the common constraints, namely,
\begin{equation}\label{com}
c_{1111}-c_{3333}>0\,,\qquad c_{1212}-c_{2323}>0\,,\qquad {\mathrm {and}}\qquad c_{3333}-c_{2323}>0\,,
\end{equation}
limit the possible values of the unknown elasticity parameters of these tensors.
That limitation occurres not to be entirely overlapping with the limitation caused by the fundamental constraints, that is, by the stability conditions.
In terms of a graphical representation of both limitations imposed on parameters---as shown in Figures~\ref{fig:one}--\ref{fig:three}---we say that it might be useful to consider the intersection of two restricted areas. 
Thus, the consideration of both the common and fundamental constraints can be helpful, for instance, in the prediction or estimation of the unknown parameters.

We notice that the less parameters of a TI tensor are known, the more useful the common constraints are.
For example, in the first case (Section~\ref{sec:caseone}), where we know the values of only $c_{1111}$ and $c_{3333}$, the aforementioned constraints significantly limit the possible values of the remaining three parameters.
This information additionally restricts the one provided by the fundamental constraints.
In the third case (Section~\ref{sec:casethree}), where we know the values of three elasticity parameters, the impact of the common constraints is diminished, as compared to the first case.
Whereas, in the last case (Section~\ref{sec:casefour}), where we only do not know the value of $c_{1133}$, the information from the common constraints is useless, since $c_{1133}$ does not appear in expression~(\ref{com}).

Apart from the amount of the unknown parameters---discussed in the paragraph above---the effect on usefulness of the common constraints varies also with their combination.  
If we compare the first case with the second one (Section~\ref{sec:casetwo}), we see that the common constraints restrict the area of possible values of elasticity parameters more significantly in the first example.
It means that the common constraints are more useful if we know the values of $c_{1111}$ and $c_{3333}$, than the values of $c_{2323}$ and $c_{3333}$.
Further, in the last case the common constraints are useless. 
However, if we consider another case, where we do not know the value of, for instance, $c_{1111}$, instead of $c_{1133}$, then these constraints might be useful due to the appearance of $c_{1111}$ in expression~(\ref{com}).

The issue with the common constraints investigated in this paper is that, in some specific situations, they might be irrelevant to the actual behaviour of seismic waves. 
If it is not the case, it might be useful to impose the aforementioned restrictions along with the stability conditions in order to better estimate the unknown elasticity parameters of a TI tensor.
\section*{Acknowledgements}
We wish to acknowledge consultations with our supervisor Michael A. Slawinski.
This research was performed in the context of The Geomechanics Project supported by Husky Energy. 
Also, this research is partially supported by the Natural Sciences and Engineering Research Council of Canada, grant 202259. 
\bibliographystyle{apa}
\bibliography{bibliography}
\end{document}